# Moored observations of turbulent mixing events in deep Lake Garda (I)

by Hans van Haren[*1], Sebastiano Piccolroaz[2,3],
Marina Amadori[4,5], Marco Toffolon[4], Henk A. Dijkstra[2]

[1]Royal Netherlands Institute for Sea Research (NIOZ) and Utrecht University, P.O. Box 59, 1790 AB Den Burg, the Netherlands.
*Corresponding author e-mail: hans.van.haren@nioz.nl

[2]Institute for Marine and Atmospheric research Utrecht (IMAU), Department of Physics, Utrecht University, Utrecht, the Netherlands.
h.a.dijkstra@uu.nl

[3]Present address: École Polytechnique Fédérale de Lausanne, Physics of Aquatic Systems Laboratory - Margaretha Kamprad Chair – APHYS GR A2 412, CH-1015 Lausanne, Switzerland.
sebastiano.piccolroaz@epfl.ch.

[4]Department of Civil, Environmental, and Mechanical Engineering, University of Trento, Italy.
marina.amadori@unitn.it; marco.toffolon@unitn.it

[5]Present address: Institute for electromagnetic sensing of the environment (IREA), National Research Council, Milan, Italy.

Keywords: Yearlong high-resolution moored temperature observations; deep Lake Garda; internal waves; turbulent overturning; convective cooling

*Abstract*

Deep water circulation and mixing processes in deep lakes are largely unknown, although they are responsible for the transport of matter, nutrients and pollutants. Such a lack of knowledge cannot be reliably provided by numerical hydrodynamic modelling studies because detailed observations are typically not available to validate them. To overcome some of these deficiencies, a dedicated yearlong mooring comprising 100 high-resolution temperature sensors and a single current meter were located in the deeper half of the 344 m deepest point of the subalpine Lake Garda (Italy). The observations show peaks and calms of turbulent exchange, besides ubiquitous internal wave activity. In late winter, northerly winds activate episodic deep convective overturning, the dense water being subsequently advected along the lake-floor. Besides deep convection, such winds also set-up seiches and inertial waves that are associated with about 100 times larger turbulence dissipation rates than that by semidiurnal internal wave breaking observed in summer. In the lower 60 m above the lake-floor however, the average turbulence dissipation rate is approximately constant in value year-around, being about 10 times larger than open-ocean values, except during deep convection episodes.

The occurrence and extent of vertical mixing is a critical factor regulating important trophic, biological, and water-quality-related processes in lakes (Salmaso et al. 2003; Berger et al. 2007; Boehrer et al. 2008; Dokulil 2014; Leoni et al. 2014). In fact, vertical mixing directly controls the fluxes of energy, oxygen, and pollutants along the water column, and the recycling of algal nutrients from deep layers to the surface. The understanding of the mixing processes of a lake is therefore a key prerequisite to characterize its response to varying external forcing and, in turn, the possible effects on its health status. In this respect, the availability of direct measurements of lake velocities and of turbulence quantities provides substantial information on mixing dynamics and is an essential condition (often not available) to properly tune the mixing schemes and validate the results of thermo-hydrodynamic lake models. In fact, verifying simulated mixing coefficients is not straightforward in many practical cases, as measurements are not always available and models might not take into account all relevant processes contributing to the turbulence field. For example, in the Delft-3D model of Lake Garda (Amadori et al. 2018), where a k-$\varepsilon$ turbulence scheme is used, the turbulent mixing is known to be inaccurate below the thermocline, due to the parameterizations of small-scale turbulent processes (e.g. internal seiches, see also Goudsmit et al. 2002 and Perroud et al. 2009 for one-dimensional applications). In this regard, observations can help identifying additional important mixing processes and lead to better parameterizations of these processes.

In confined enclosed basins, mixing in the interior as well as at the sediment boundary is largely determined by internal waves, thanks to their important role in linking large-scale motion triggered by external forcing to small-scale turbulent structures (e.g., Imboden and Wüest 1995; Wüest and Lorke 2003) and to the interaction with sloping boundaries (Lorrai et al., 2011). Internal waves are abundant in a lake interior, anytime and everywhere, as has been demonstrated in various observational studies in different lakes (e.g., Farmer 1978; Thorpe et al. 1996; Boegman et al. 2003; Preusse 2012). They can be generated by a sudden variation in environmental conditions, like a variation in current-flow over lake-floor-topography, by geostrophic adjustment in response to the passage of, e.g., an atmospheric disturbance, or in larger lakes by interaction of gravitational (semidiurnal) internal tides with underwater topography, similar to the ocean interior (Gill 1982). However, the most important factor triggering internal waves in lakes is certainly the wind forcing, which continuously provides space and time-varying perturbations in the shear stress at the surface (Antenucci and Imberger 2003; Lemmin et al. 2005; Valerio et al. 2012), thus exciting the internal baroclinic response of the lake as a closed basin of finite length and given topography (Wang et al. 2000; Wüest and Lorke 2003; Lorke 2007; Valerio et al. 2019). In this regard, an important internal mode of variability is represented by standing internal seiches (e.g., Lemmin et al. 2005). Such waves are generated by the wind-driven set-up of the water surface at the leeward end of the lake and contextual upward tilting of deep layers at the windward end of the lake, which starts to oscillate once the wind relaxes. Overall, the internal-wave field in a lake has a basin-particular resonance response, but is also chiefly modulated by the wind forcing periodicity.

Internal waves are supported by vertical density ($\rho$) stratification that is mainly a result of the solar heating from above, warm water being less dense than cold water (e.g., LeBlond and Mysak 1978). Freely propagating gravitational waves span a wide range of frequencies, between inertial (low) frequency $f = 2\Omega\sin\varphi$, where $\Omega$ denotes the Earth rotational vector and $\varphi$ the latitude, and stratification-related buoyancy (high) frequency $N = \{-g/\rho(d\rho/dz + g\rho/c_s^2)\}^{1/2}$ (e.g., Gill 1982), where g denotes the acceleration of gravity and $c_s$ the speed of sound reflecting the thermobaric correction due to pressure-compressibility effects. As internal waves may locally deform, 'straining' the stratification thereby intermittently forming thinner more stratified and thicker more homogeneous layers, a 'background' N is computed over a typical vertical length-scale of $\Delta z = 100$ m, which roughly represents the scale of the larger internal waves. Additionally, such N should also be computed over an associated time-scale, evaluated at least as an average over the buoyancy scale if known from high-frequency internal waves, but preferably equal to the inertial time-scale of half a pendulum day. Over such time-scales, turbulent overturning such as by internal wave breaking is expected to be averaged out.

Thus far, vertical turbulent mixing has been mainly observed using shipborne vertical profiling instrumentation in deep lakes. In a comprehensive study of the weakly stratified interior of Lake Baikal Ravens et al. (2000) observed mean $N = 1.4\times10^{-4}$ s$^{-1}$, mean dissipation rates $O(10^{-10})$ m$^2$ s$^{-3}$) and mean eddy diffusivities $1-9\times10^{-3}$ m$^2$ s$^{-1}$. From their microstructure profiles and CTD data they established a mean mixing efficiency of $0.16\pm0.1$, which they found consistent with the mean value of 0.2 in more stratified natural waters (e.g. Osborn 1980; Oakey 1982). Recently, Portwood et al. (2019) analyzed direct numerical simulations estimating a mixing coefficient near the classical bound of 0.2 in stationary flows for a wide range of conditions. Ravens et al. (2000) also noted an important contribution of turbulent kinetic energy production in the boundary layer above the lake floor, which was about 10% lower than the energy found in the lake interior. The lake interior turbulence was mainly driven by wind-induced inertial motions.

In this paper, we report on direct measurements of deep interior mixing acquired by a yearlong moored station installed in the bottom 150 m near the deepest point of the subalpine Lake Garda (Italy). Lake Garda's interior is weakly stratified, but its mean N is about five times larger than Lake Baikal's interior value. High-resolution vertical temperature profiles and single point flow velocity observations are used to investigate the generation of vertical mixing due to various quasi-periodic motions, throughout the seasons and in relation to different processes such as internal waves and buoyancy driven deep convection in wintertime. As in Lake Baikal, we hypothesize that the main driver for interior vertical turbulent mixing is wind-induced internal wave breaking. We want to compare such mixing with that by deep convection.

## *Materials and methods*

**Study site**

Lake Garda is the largest lake in Italy and a popular attraction with a few million tourists visiting it every year. However, from a hydrodynamic perspective it is also one of the most unexplored large Alpine lakes (Toffolon et al. 2017; Amadori et al. 2018). It has a maximum depth of 346 m and a mean depth of 133 m over a heterogeneous bathymetry (see Fig. 1a), which is characterized by an elongated, narrow, and deep basin to the North and a round shape and shallower basin to the South. Lake Garda is classified as an oligomictic lake (Wetzel 2001), where variable periods of incomplete mixing are interspersed with occasional events of complete mixing during cold winters, similarly to the other major deep lakes located just south of the Alps (Ambrosetti and Barbanti 1999; Salmaso 2005). While deep mixing events are traditionally associated to buoyancy-driven convection during cold winters, recent in-situ observations and modelling results proved that wind-driven advection and secondary flows influenced by the Coriolis acceleration appreciably contribute to deep ventilation processes (Piccolroaz et al. 2019).

Starting in 2017, substantial efforts have been spent to collect high quality turbulence and hydrodynamic measurements in Lake Garda. During 2017-2018, turbulent mixing in the upper 100 m has been measured at several locations in the lake using a microstructure turbulence profiler (Piccolroaz et al. 2019; MicroCTD manufactured by Rockland Scientific International). In addition, between May 2017 and May 2018 an instrumented mooring line constituted of a string of high-resolution temperature sensors and a current meter was operating in the lake. Here we present the analysis of the deep interior mixing measurements acquired by this moored instrumentation.

**Operational site characteristics**

The taut-wire sub-surface mooring was deployed near the deepest point of Lake Garda, off the town of Brenzone (Fig. 1). The geographic location of the mooring was: latitude $\varphi = 45°$ 42.947´N ($f = 1.04 \times 10^{-4}$ s$^{-1}$, so that the inertial period is 16.7 h), longitude 10° 44.567´E. The instrumentation operated for one year, between 10 UTC on 24/05/17 (yearday 144.42) and 09 UTC on 31/05/18 (yearday 516.37; days in 2018 are +365). Locally, the water depth was H = 344 m with a relatively weak bottom slope (100 m scale) of about $\beta = 3°$ (based on the lake bathymetry realized by the Hydrographic Institute of the Italian Navy in 1967, see Fig 1a). Some 300 m to the East and 1.3 km to the West, the lake-floor shallows with slopes varying between 10° and 20°.

The relatively flat central part and steep sides of the underwater topography generate and reflect internal waves differently. In contrast with surface (wind) waves, internal waves can freely propagate in three spatial dimensions with an angle $\gamma(\sigma, N) = \arcsin[((\sigma^2-f^2)/(N^2-f^2))^{1/2}]$ (e.g., LeBlond and Mysak 1978) from the vertical, which depends on wave frequency $\sigma$ and local N $\approx 7 \times 10^{-4}$ s$^{-1}$ at given f. For deep Lake Garda, the average semidiurnal solar (S$_2$, 12.0 hours period) angle is $\gamma(S_2, N) \approx 8°$ and the mooring-site bottom slope $\beta < \gamma(S_2, N)$ is sub-critical for these internal waves and about equal ('critical') for near-inertial waves $\beta \approx \gamma(1.1f, N)$, for given N. Conversely, the nearby steeper sloping sides of the basin are generally super-critical for S$_2$-waves, and thus for near-inertial waves. Slope-criticality implies a condition where internal waves may trap their energy at a topographic site of the same slope, while they reflect back into the lake interior at super-critical slopes or reflect forward towards the shelf in the case of sub-critical slope. Previous results suggest substantial internal wave breaking and turbulence generation above uniform critical slopes (e.g., Eriksen 1982; Dauxois et al. 2004; Lorke et al. 2006), but recent modelling and ocean-observational efforts show more intense internal wave breaking turbulence above super-critical slopes (van Haren et al. 2015; Winters 2015; Sarkar and Scotti 2017).

It is noted that lakes are not rigid environments in which the density stratification and topographic slopes are fixed. In reality, topographic slopes vary in value over many length-scales and $N(x,y,z,t)$ is not constant but may vary considerably with space and time, e.g., upon sub-inertial modifications like warming from above, wind-induced flows and seiches. As a result, internal wave slope-criticality is seldom found over any prolonged period of time. In Lake Garda large-scale N varies between about $10^{-4}$ (~f) and $10^{-3}$ s$^{-1}$ in the deeper half. Even under the strongest mean large-scale stratification, the local bottom-slope at the mooring site will be sub-critical for the $S_2$-wave slope. On small (~2 m) scales in thin layers due to wave straining, stratification may increase to $N_{max} = 4\times10^{-3}$ s$^{-1}$, which can extend the internal wave band to higher frequencies locally.

**Mooring characteristics**

As one cannot use (large) research vessels on Lake Garda, the mooring top-buoy was towed to the dedicated position with a fast boat thanks to the support of the firefighters nautical rescue team of the Autonomous Province of Trento (Fig. 1b).

The mooring consisted of floatation providing 1100 N net buoyancy near its top at about 150 m below the lake-surface. Due to shipping, fishing, and environmental constraints, the unattended instrumented line could not be closer to the surface. Below the floatation, a single point Nortek AquaDopp current meter was at 157 meters above the bottom (corresponding to 187 m below the water surface; henceforth we will refer to depths in terms of mab, meaning "meters above the bottom"). The AquaDopp sampled current components [u v w] and acoustic echo intensity I, storing ensemble data every 300 s. Based on these data the (horizontal) kinetic energy KE is defined as KE = $(u^2 + v^2)$. Its tilt and pressure sensors demonstrated that the top-buoys did not move more than 0.01/0.05 m vertically and not more than 1.5/5 m horizontally, under general/maximum current speeds of 0.05/0.17 m s$^{-1}$, respectively.

Between 6 and 154.5 mab, 100 'NIOZ4' self-contained high-resolution temperature (T) sensors were deployed at 1.5 m vertical intervals, sampling for 0.12 s at a rate of once per 2 s, with a precision of better than 0.5 mK after drift correction, and a noise level of less than 0.1 mK (van Haren 2018). The T-sensors' internal clocks are synchronized via induction every 4 hours. Thus, each entire vertical profile of 148.5 m is measured within a clock-accuracy of less than 0.02 s, which is not achievable by shipborne profiling instrumentation that would take more than 150 s to cover such distance (with a typical profiling speed of <1 m s$^{-1}$). As a result, the amount of statistical data provided by the moored temperature sensors is much larger than achievable using standard shipborne or wire-walker profilers.

Drift of nominally 1 mK/mo for aged sensors is corrected by fitting typically four-day mean temperature profiles to a smooth statically stable multiple order polynomial profile. Up to yearday 350, 7 months after deployment, 7 sensors showed various electronic (noise, calibration) problems and are not further considered. Their data are linearly interpolated. At yearday 450, 40 sensors had failed due to inferior batteries. For data until then interpolation is still possible, although turbulence values (see below) are biased low by about 30%. For data after yearday 470, when 50% of sensors had failed, no further analyses of turbulence characteristics were performed.

**Additional observations**

The moored observations are supported by weather data from meteorological stations about 10 km to the north of the mooring site (station 'Limone-sul-Garda') and about 10 km to the south of mooring (station 'Toscolano-Maderno'), both operated by the Environmental Protection Agency of Regione Lombardia ARPAL (see Fig. 1a).

Further support is provided by multiple SBE 19plus V2 SeaCAT Conductivity Temperature Depth CTD-profiles that were measured by the Environmental Protection Agency of Regione Veneto ARPAV within 1 km from the mooring site over almost the entire water depth eight times, roughly distributed over the year-long observational period. The CTD-data are used in fine-tuning the correction for instrumental drift of the temperature sensors. For the Lake Garda data, they are especially important for the proper establishment of the polynomial correction-profile in the lower 30 m above the lake-floor. The CTD-data are also used to establish a, preferably linear, temperature-density relationship, which is more straightforward for data from a lake than from the ocean where salinity may significantly contribute to density variations. A tight temperature-density relationship enables the computation of turbulence values from the moored T-sensor observations. In order to do so, the moored T-sensor data are first transferred to potential temperature ($\theta$) values (IOC et al. 2010) to correct for pressure-compressibility effects. To account for local chemical ion-composition, salinity is computed following the parametrization by Salmaso and Decet (1998). Salinity contributes about 15% to density variations that are thus dominated by temperature variations in the deeper half of Lake Garda. Over the range of moored temperature observations, the relationship between potential density anomaly referenced to the surface ($\sigma_\theta$) and $\theta$ is obtained from CTD-data (shown in Fig. 2d), $\sigma_\theta = \alpha\theta$, where $\alpha = -0.1129 \pm 0.0003$ kg m$^{-3}$°C$^{-1}$ denotes the regression coefficient under local conditions over the range of T-sensors. This regression coefficient remains constant to within about 5% for most of the mooring period. When deep dense water passes after day 428 (March 5, 2018), it changes to $\alpha = -0.054 \pm 0.0007$ kg m$^{-3}$°C$^{-1}$, still relatively tight.

**Turbulence quantification**

We obtained turbulence values using the moored temperature sensor data as tracers for density variations by calculating 'overturning' scales. These scales followed after reordering (sorting) every 2 s the 148.5 m high $\Theta$-profile, which may contain inversions, into a stable monotonic profile without inversions following Thorpe (1977), who proposed the method for shipborne CTD-data (in a lake). After comparing observed and reordered profiles, displacements (d) were calculated necessary for generating the reordered stable profile. Certain tests apply to disregard apparent displacements associated with instrumental noise and post-calibration errors (Galbraith and Kelley 1996). Such a test threshold is very low for rigorously moored NIOZ-temperature sensor data, $<5\times10^{-4}$°C (van Haren 2018). Then the turbulence dissipation rate, with the same dimensional units as the vertical turbulent flux, reads (Thorpe 1977),

$$\varepsilon = 0.64d^2N^3 \qquad [m^2\,s^{-3}], \tag{1}$$

where N denotes the buoyancy frequency computed from each of the reordered, essentially statically stable, vertical density profiles. The numerical constant follows from empirically relating the rms displacement or Thorpe overturning scale with the Ozmidov-scale $L_O$ of stratified turbulence $c_1 = |L_O/d|_{rms} = 0.8$ (Dillon 1982). As an additional vertical turbulence parameter, diffusivity $K_z = \Gamma\varepsilon N^{-2}$ can be computed. Hereby, a constant mixing efficiency is used of $\Gamma = 0.2$ (Osborn 1980; Oakey 1982; Gregg et al. 2018; Portwood et al. 2019) as a mean value, in a large data-range scattered over a factor of ten, for the conversion of kinetic into potential energy, so that,

$$K_z = 0.128d^2N \qquad [m^2\,s^{-1}]. \tag{2}$$

It has been argued that $c_1$ and $\Gamma$ vary as a function of stratification, stage and type of turbulence generation (e.g., Chamalla and Sarkar 2013; Bouffard and Boegman 2013; Mater et al. 2015; Garanaik and Venayamoorthy 2019). While this is acknowledged for specific unique conditions like pure convection, low Reynolds number flows, there are several reasons why it may not apply, and cannot be applied, to the present data, as partially outlined in van Haren (2017). The moored high-resolution temperature sensor provide a priori insight in particular internal wave turbulence processes, but we are not primarily concerned with individual overturning values. Instead we give "suitably averaged" turbulence values, as detailed below. In any given high Reynolds number environment like natural waters, shear- and buoyancy-driven turbulence intermingle and are hard to separate. Examples of numerical modelling studies of particular stages of instability developments show that a finger of convective instability develops secondary shear instability along its fringe (Li and Li 2006), while the roll-up stage of shear instability develops secondary convective instability mushrooms (Matsumoto and Hoshino 2004). Comparison between calculated turbulence values using shear measurements and using Thorpe overturning scales with $c_1 = 0.8$ from areas with such mixtures of turbulence development above sloping ocean topography led to 'consistent results' (Nash et al. 2007) and 'results similar to within a factor of three' (van Haren and Gostiaux 2012). A factor of two is the minimum error range for turbulence parameter values (Oakey 1982). It is presently impossible to try better in natural waters where turbulence values from individual overturns vary over four orders of magnitude (e.g., Oakey 1982; van Haren and Gostiaux 2012; Gregg et al. 2018).

Thus, from the argumentation above and the reasoning in Mater et al. (2015), internal wave breaking unlikely biases turbulence dissipation rates computed from Thorpe overturning scales by more than a factor of two-three, provided some suitable time-space averaging is done instead of considering single profiles. This is within the range of our error. Likewise, Ravens et al. (2000) found a mixing efficiency very close to 0.2 in very weakly stratified deep interior Lake Baikal.

As for averaging the present data, we first calculated non-averaged d in (1) and (2) for high-resolution profiles of $K_z(z, t)$ and $\varepsilon(z, t)$. Subsequently, we calculated 'mean' turbulence values by averaging the parameters in the vertical [ ] or in time <>, or both. This averaging was a straight arithmetic average for dissipation rate, but for eddy diffusivity the turbulent flux was averaged first before obtaining mean-$K_z$ and for the buoyancy frequency the stratification was averaged first before obtaining mean-N. Averaging over at least the largest overturning scale is required (Thorpe 1977). The rapid sampling of vertical profiles of moored high-resolution T-sensor data warrants sufficient samples for averaging over all different turbulence characteristics. A buoyancy period of 3 h averages 5000 profiles, an inertial period averages 30,000 profiles, a four-day period averages 160,000 profiles.

The errors in the mean turbulence parameter estimates thus obtained depend on the displacement estimates, the variation in turbulence type and age, the error in N and the error in the temperature-density relationship, while the instrumental noise error of the T-sensors is negligible.

*Observations*

**CTD-profile over the entire water-depth**

The water temperature in Lake Garda is well above 4°C, so that the density-temperature relationship is monotonic all year round. The stable stratification due to the heating of the lake surface water is deformed in layers, with multiple diurnal and seasonal thermoclines still prominent in individual profiles also in fall and early winter (see the shipborne CTD example from mid-November in Fig. 2a-c). As is clear from the vertical profiles, salinity contributes positively to the density stratification, but its contribution to density variations is relatively weak, around a factor of 10 less, compared to that of temperature, also in deep waters (z<-150 m). The 13 March 2018 profile is unique with enhanced temperature and salinity decreases with depth near the lake-floor. These decreases are associated with winter deep dense water formation. During this episode, the temperature-density relationship changes (Fig. 2d). Even in the late autumn, in the upper half of the water column (z>-150 m), stratification is relatively strong in particular layers. However, it does not imply that turbulent exchange cannot occur and vertical fluxes of, e.g., nutrients are not necessarily negligible, as will become clear in examples below when the layers go deeper upon mixing from the surface down. In the lower half of the water column (z<-150 m), temperature variations seem very small, but the waters are not permanently homogeneous and the stratification supports internal waves that may also break, as will be demonstrated from the detailed observations presented forthwith.

**Entire time series**

Lake Garda is characterized by along-lake regular (especially in warm and sunny days) daily alternate winds in the northern elongated part, which generally are stronger and more persistent when blowing from the north (to the south) than vice-versa (Fig. 3a). Strong wind episodes occur year-round, on a periodic daily basis (according to the daily alternation between the Peler and Ora del Garda breezes during the warm season, Giovannini et al. 2015) and on a non-periodic approximately 10-day basis (in correspondence with episodic northerly Föhn winds, particularly abundant in the cold season, Piccolroaz et al. 2019). The wind events directly affect near-surface flow-currents, while the water-flow around mid-depth (Fig. 3b) is inversely correlated with these wind events (as will become clearer in detailed images below). Mid-depth flow-speeds peak up to 0.15 m s$^{-1}$ upon an early winter wind event around day 318. Otherwise mid-depth flow speeds are typically about 0.05 m s$^{-1}$, with several days periodicities associated with basin responses of standing waves (seiches) to wind events, and with shorter periodicities associated with diurnal variations, inertial motions and internal waves having periods between inertial and buoyancy. For example, seiches of approximately two days periodicity occur for the duration of about 3 to 4 periods between days 254 and 261 (late summer, strongly stratified period). Seiches with approximately five days periodicity occur between days 360 and 380 (winter, weakly stratified period). These periodicities correspond with those found in the Delft-3D numerical model of Lake Garda (Amadori et al. 2018).

Over most of spring to late autumn, the air temperature is higher than the mid-depth ($z = -190$ m) water temperature (Fig. 3c). In late autumn from about day 325 onward, air temperature drops episodically below the mid-depth water temperature. Following air temperature cooling, the lake surface temperatures drops and the lake starts destratifying in this season determining unstable situations at the surface, with contextual deepening of the epilimnion and enhancement of mixing. This is detected by the T-sensor string in deep water $z<-190$ m, where heat is observed to be pushed down over time (Fig. 3d). This already occurs in summer, weakly, and progressively more in winter, where heating reaches down to the lake-floor.

Only upon a large cooling event in late winter occurred around day 425 with the passage of a cold Siberian front (named "Burian"), sudden and efficient cooling of near-bottom waters is observed due to deep buoyancy-driven convection. Such an event is characterized by large turbulent overturning resulting in commensurate turbulence dissipation rates (Fig. 3e). Before this event, episodes of relatively enhanced, certainly non-negligible, turbulence are observed that are associated with northerly wind events and/or with internal wave breaking, as will be demonstrated in detailed examples below. Some local enhancement of turbulence dissipation rate seems associated with the two different seiche-periods mentioned above and with the strong northerly wind event. Small-scale internal waves are not discernable in the overview plot of Fig. 3d, but will be visible in detailed plots below.

Averaged over the 148.5 m vertical range (indicated by the brackets [ ]) and the 300 days of observations (indicated by the brackets < >) of Fig. 3, the mean turbulence dissipation rate is $<[\varepsilon]> = 8\pm5\times10^{-9}$ m$^2$ s$^{-3}$, while $<[N]> = 7.7\pm1\times10^{-4}$ s$^{-1}$ (cf. Table 1). For the period up to day 425 before the sudden cooling $<[\varepsilon]> = 1.1\pm0.6\times10^{-9}$ m$^2$ s$^{-3}$, while for the period from day 425 to 429 $<[\varepsilon]> = 4\pm3\times10^{-8}$ m$^2$ s$^{-3}$. The turbulence values up to day 425 are comparable with, albeit up to half an order of magnitude larger than, open-ocean values (e.g., Gregg 1989; Polzin et al. 1997; van Haren 2019) and other lake interior observations (e.g., Goudsmit et al. 1997; Wüest and Lorke 2003).

The above periodicities stand out more clearly in the average spectra from mid-depth (z = -187 m) current meter observations (Fig. 4a). The spectra of different parameters show different characteristics when compared with the meaningful harmonic frequencies: diurnal ($S_1$), semidiurnal ($S_2$) and a quarter of day ($S_4$), inertial (f) and buoyancy (N). The pressure record, representing the lake-surface barotropic variations, have a broad band of variations on a yearly basis, with about 1 m higher levels in spring compared to late summer/autumn, due to the different seasonality between rainfall, snowmelt, hydropower releases and use of lake water for irrigation purposes. Pressure variations also have a periodicity of 10 days and more (frequency <0.1 cpd, short for 'cycles per day'), besides narrow-band quasi-deterministic variations at solar diurnal and, slightly less energetic, semidiurnal frequencies that are related with wind alternation. The kinetic energy of current flows varies over a broad band of 2 to 5 days periodicities, over narrow bands peaking around solar diurnal and semidiurnal frequencies and over a moderately broad band at a frequency just above the inertial frequency. The latter is typical for (near-inertial) internal wave motions. The mid-depth (z = -187 m) current meter temperature shows hardly any significant peaks, but weak enhancements around 10 days periodicity, roughly around 0.3 cpd, just above f including a small peak at the solar semidiurnal frequency and around 2f. The spectral slope becomes steeper at just above the buoyancy frequency.

The average (up to day 425) temperature sensor spectra at 5 different levels from the lower half of the water column show a decrease in variance by more than one order of magnitude towards the lake-floor (Fig. 4b). In this figure panel, the plotted spectra are scaled with a -5/3-power law in frequency, which represents the common inertial subrange of shear-induced turbulence decay with frequency for a passive scalar (Tennekes and Lumley 1972; Warhaft 2000). Relative to this spectral slope (on a log-log plot), it is observed that the entire internal wave band is dominated by turbulent motions around mid-depth (z = -191 m), rather than following the canonical slope of $\sigma^{-2}$ representing internal waves (Garrett and Munk 1972) or thermal fine structure advected passed the sensors (Phillips 1971). Internal wave slopes of $\sigma^{-2}$ do occur in spectra from deeper down (z < -262 m), which show a rather broad peak at semidiurnal frequency and a progressively narrowing internal wave band with increasing depth and decreasing stratification. All spectra roll-off towards lower (sub-inertial) frequencies at about 1.1f, but less clearly towards the lake-floor. The frequency of 1.1f either reflect near-inertial waves that propagate from a distant source northward of the observational site, which however cannot be relevant in Lake Garda because of the small size of the lake compared with the required latitudinal variation, or the addition to planetary vorticity of local vorticity, e.g., due to sub-inertial seiche or other motions. The statistically non-significant sub-inertial extent of the spectral peak at great depths suggests influences of the non-traditional horizontal component of the Coriolis force that expand the internal wave band to <f and >N-frequencies and that become dynamically important in weak stratification only (LeBlond and Mysak 1978). Diurnal motions are not significantly peaking in any of the temperature spectra.

In the following sections, we examine some specific processes that can be noticed in the different periods, when the variability is related both to the external forcing (e.g., wind forcing, deep convective cooling) and to the strength and vertical distribution of stratification in the lake.

**Wintertime northerly wind events**

A four-day magnification of data shows the sudden onset of winds from the north and their response in the lower half of the lake-center (Fig. 5). At both the northern and southern meteorological stations, the moderate southerly winds change to strong northerly winds (blowing to the south and reaching more than 10 m s$^{-1}$) simultaneously within an hour or so (Fig. 5a). The mid-depth water-current initially responds with flow to the north building up within about 6 h, opposite to the wind and presumably in shear with water-flow near the surface, before turning to the south about 17 h later (Fig. 5b). This corresponds with the 16.7 h local inertial period. Upon the southerly flow, temperature increases at nearly all depths investigated. With it, high-frequency internal waves are initiated that are supported by local stratification that is somewhat enhanced due to straining (Fig. 5c). The stratification varies down to N ~ f in the lower 10 to 60 m above the lake-floor, with the notion that the extent of this 'layer' varies considerably over time with the high-frequency internal waves.

The internal waves episodically break, increasing local turbulence, foremost in weaker stratified layers in between thin stratified layers in the interior, e.g. around z = -200 m between days 317.5 and 318, but notably also in the lower 30 m above the lake-floor (Fig. 5d). The largest turbulent patch near the bottom is associated with the turn to southerly flow around mid-depth and the deepening of the stratification around day 316.8. The wind event and internal wave breaking thus generate near-bottom turbulence. For these 148.5 m thick layer and four days of observations, the average turbulence values are <[ε]> = 6±4×10$^{-10}$ m$^2$ s$^{-3}$ and <[K$_z$]> = 2±1×10$^{-4}$ m$^2$ s$^{-1}$, while <[N]> = 7.6±1×10$^{-4}$ s$^{-1}$. These values compare with open-ocean and open-lake values away from (sloping) boundaries near the surface and sea-floor (e.g., Gregg 1989; Polzin et al. 1997; Ravens et al. 2000; Wüest and Lorke 2003; van Haren 2019). Molecular diffusion of heat has values of about 10$^{-7}$ m$^2$ s$^{-1}$ in water.

Details of this interior and near-bottom internal wave breaking induced mixing are visible in magnifications of 8, 4 and 1 h panels (Fig. 6). The anecdotic images show that the shape of isotherms is far from sinusoidal (i.e. single frequency linear internal waves), but rather it is skewed and highly nonlinear, due to episodic instabilities and associated overturning. The amplitudes and overturns are not small, and extend several tens of meters in the vertical. The synchronization between the upper stratified layer around z = -220 m, characterized by more roll-up overturning, and isotherm displacements closer to the lake-floor, characterized by more convection-like up- and down-turbulent movements, is irregular. A few examples are indicated by red ellipses in Fig. 6. Larger than 10 m overturns do occur episodically and are not a rare phenomenon.

The apparent lack of correlation between interior and near-bottom motions implies that internal waves do not occur as local vertical mode-1 but mostly as higher local vertical modes of isotherms oscillating 180° out-of-phase, likely associated with local overturning.

**Late winter cold water influx**

At the end of winter after about two months with cold air temperature determining the deepening of the surface well mixed layer through episodes of near-surface cooling (Fig. 3c), a drastic cooling is observed in the moored temperature data (Fig. 7). Under circumstances with weak and cyclic winds to the north and associated weak but persistent mid-depth currents to the south and weakly stratified waters through the array of temperature sensors on days 426 and 427, temperature falls abruptly by <-0.1 °C on day 428.0, near the lake-floor first. Whilst it associates with a deep cooling (and freshening, Fig. 2) event, the cooling is not effective locally, but apparently occurred to the north of the mooring (according to the southward deep current, Fig. 7b). Subsequently, relatively cool waters are advected underneath warmer waters above, which generates a statically unstable situation with a natural convectively driven overturning that appears as partial shear-driven turbulence in the interior. This unstable episode follows a local natural convective instability of approximately the same size of half a day of turbulent overturning between days 427.1 and 427.6. It cannot be established whether this cooling arrived from the lake-surface, or from the sides of the basin around mid-depth.

For these 148.5 m thick layer and four days of observations, the average turbulence values are $<[\varepsilon]> = 6\pm4\times10^{-8}$ m$^2$ s$^{-3}$ and $<[K_z]> = 3\pm1.2\times10^{-2}$ m$^2$ s$^{-1}$, while $<[N]> = 6.3\pm1\times10^{-4}$ s$^{-1}$. These values are two orders of magnitude larger than during the wintertime northerly wind events (Fig. 5). They compare with the upper limit of values observed in the interior of Lake Baikal (Ravens et al. 2000) and in internal wave breaking above steep deep-ocean sloping topography (van Haren and Gostiaux 2012).

**Late spring/early summer internal waves**

In late spring/early summer, when stratification is well-established throughout most of the lake, the interior is far from quiescent (Fig. 8). Internal waves are observed that have amplitudes of typically 10 m, with crest-trough values larger than 40 m episodically (Fig. 8 c,d). Vertical phases and isothermal straining vary rapidly, so that vertical local low-mode internal waves are nearly always accompanied by higher modes. It is noted that only half of the water column is monitored so that a single vertical mode-1 cannot be well established. The dominant periodicity is semidiurnal, but distributed over a rather broad band of frequencies because the internal waves occur in groups of 5-7 waves, similar to surface wind-wave groups. Such internal wave group intermittency is attributed to multiple wave-wave interactions and variations in the background stratification that supports the waves. The semidiurnal variation is expected to be generated as a first harmonic of the diurnal wind variation (Fig. 8a), whose speed is not very strong so that (semi)diurnal current-flows are also rather weak of a few 0.01 m s$^{-1}$. The associated turbulence is mainly concentrated in the lower 70 m above the lake-floor, with some intense semidiurnal pulses up to 40 m above the lake-floor, but otherwise rather weak in the interior (Fig. 8d). Apparently the (semidiurnal) internal wave interior shear does not induce strong or large overturning locally in the interior, thus turbulence is relatively weak.

For these 148.5 m thick layer and four days of observations, the average turbulence values are <[$\varepsilon$]> = 7±4×10$^{-11}$ m$^2$ s$^{-3}$ and <[$K_z$]> = 3±1.5×10$^{-5}$ m$^2$ s$^{-1}$, while <[N]> = 6.8±1×10$^{-4}$ s$^{-1}$. These values are half an order of magnitude smaller than during the wintertime northerly wind events (Fig. 5). These values are compatible with those found in the hypolimnion of deep lakes during the stratified period and quiescent episodes (e.g., Wüest and Lorke 2003).

**Late winter inertial waves**

In contrast with the previous example, in late winter/early spring the interior overturning can be very strong while the direct effects on the lake-floor remain rather small (Fig. 9). Upon a relatively strong diurnal/inertial wind oscillation, the basin responds with a group of 7-9 near-inertial waves, which commonly have a short vertical scale or relatively large vertical current difference 'shear' (LeBlond and Mysak 1978; van Haren et al. 1999). Across a given layer, which can be near-homogeneous for more than 50 m, equivalently large overturns are observed with a duration of the local buoyancy period individually and the local inertial period for the group of overturns (Fig. 9d). During the passage of the inertial waves peaking around day 446, the isotherms are most squeezed upon a downward isotherm motion from above and upward motion from the lake-floor.

For these 148.5 m thick layer and ten days of observations, the average turbulence values are $<[\varepsilon]> = 3\pm2\times10^{-9}$ m$^2$ s$^{-3}$ and $<[K_z]> = 1.3\pm0.7\times10^{-3}$ m$^2$ s$^{-1}$, while $<[N]> = 7.6\pm1\times10^{-4}$ s$^{-1}$. These values are half an order of magnitude larger than during the wintertime northerly wind events (Fig. 5). They support the hypothesis that inertial motions dominate lake-interior mixing, together with episodic deep dense water formation.

**Vertical profiles of four day mean values**

The time-average 4-day (and 10-day in the last case) vertical profiles of turbulence values for the events discussed above show considerable differences (Fig. 10). Values of turbulence dissipation rate vary over about four orders of magnitude (Fig. 10a), those of diffusivity over more than two orders of magnitude (Fig. 10b). The stratification is generally larger away from the lake-floor, but episodes occur when N has a maximum at 10 to 20 m from the lake-floor (or perhaps even closer). The effect is a maximum in turbulence dissipation rate (~flux) at some distance from the lake-floor which varies between 6 m (or even closer) to about 100 mab. Between three of the four selected periods, the profiles show similarity in the lower 60 mab, while being dissimilar higher-up. With the exception of the convective cooling period of Fig. 7 with much higher turbulence values, mean (shear-induced) turbulence values are thus approximately 'constant' year-around in this layer and about 10 times larger than open-ocean values. Conversely, Lake Garda interior (between 60 and 150 mab) mean turbulence values can be between ten times smaller and ten times larger.

*Discussion and conclusions*

The presented overview of observations on mixing processes in the deeper part of Lake Garda demonstrates variations at different scales. In general, year-long mean turbulence intensity values are comparable with open-ocean and deep lakes-interior values (e.g., Ravens et al. 2000; Wüest and Lorke 2003). Most intense turbulent overturning is associated with deep convective overturning in late winter, albeit probably not locally generated but being advected via the sloping sides of the basin, likely as cascading from the nearshore as previously observed in Lake Geneva (Fer et al. 2002) and possibly enhanced by wind-driven along-lake and lateral flows (Piccolroaz et al. 2019). Strong turbulence exceeding the mean values is also observed during a period with relatively intense near-inertial oscillations that, apparently, induce shear-induced turbulent overturning across a 50 m high near-homogeneous layer bordered by relatively large stratification in thin layers. It is known that the relatively short vertical scales of near-inertial motions can drive stratification to marginal stability which is adequate for the vertical fluxes of nutrients in seasonally stratified shallow seas like the North Sea (van Haren et al. 1999). It is more difficult to understand why the observed near-inertial induced turbulence, occurring in association with high-frequency internal wave breaking, is not affecting the lake-floor of Lake Garda.

Other internal-wave-induced turbulent overturning, albeit less intense, does reach the lake-floor. Summertime semidiurnal periodic overturning reaches up to 60 mab, and, although the observations start at 6 mab, may well resuspend material from the lake-floor. Whilst the semidiurnal motions cannot be directly gravitationally driven internal tides, they seem to be associated with the semidiurnal wind variations. As the about 0.01 m s$^{-1}$ mid-depth semidiurnal flow amplitudes unlikely generate 60 m high frictional boundary layers, which are about 1 mab for such speeds (Ekman 1905), the semidiurnal internal waves are expected to be responsible for the observed large overturning. As internal waves have relatively large vertical scales compared to inertial motions, and thus have relatively weak shear, the overturning may be partially convectively driven in the near-homogeneous deep layers. However, (secondary) shear overturning will be associated with the convection.

As the semidiurnal response is suggested to be generated following diurnal atmospheric variations, lower-frequency wind events also affect the near lake-floor turbulent motions. Via the generation of a seiche, packets of high-frequency internal waves are generated. The nonlinear waves' oscillation modifies the stratification in the lower 50 mab, just like with the semidiurnal motions, with the most intense near-lake-floor turbulent overturning occurring at the front of a group of internal waves, about one inertial period after the onset of the wind.

Internal waves are thus found to be an important medium in transferring momentum of larger-scale atmospheric-driven disturbances to the turbulent motions in the deep lake interior, foremost close to the lake-floor. The sloping sides of the lake are supercritical for most of the dominant internal wave frequencies and are expected to be important for the waves' nonlinear deformation. A quasi-permanent turbulence layer is thus established, which is only exceeded in intensity by turbulence due to rare deep cooling events and due to strong inertial wave action. This is important information to verify the mixing schemes in lake models. For example, the effects of local (e.g., seiche) vorticity on internal wave induced turbulent overturning is currently not taken into account in the Delft-3D k-$\varepsilon$ turbulence scheme, and numerical experiments have shown that turbulence generated at the bottom is too large compared to observations. Hence, the observation of deep turbulent mixing events provided here will contribute to the improvement of the turbulence parameterization as well as of the bottom friction parameterization, by identifying the key physical processes contributing to the turbulence field and the appropriate vertical resolution near the bottom to properly simulate them.

**Acknowledgments**

We greatly appreciate the assistance of the Nautical Rescue Team of 'Vigili del Fuoco-Trento' and M. van Haren around and during the deployment and recovery of the mooring. We thank M. Laan (NIOZ) for his ever-lasting temperature sensor efforts, and F. Cassano and F. Lanzillo (IMAU) for assisting on the preparation of the temperature string. This project was funded by the Faculty of Science of Utrecht University through a grant to HD and was supported by the National Marine Facilities (NIOZ) and the DICAM (University of Trento). The authors declare no conflict of interest.

**Table 1**. Time-depth mean values for turbulence dissipation rate, eddy diffusivity and buoyancy frequency computed for data of indicated figures. Values for the overview Figure 3 are split in two, including the cold period and up to the cold period.

| *Figure* | $<[\varepsilon]>$ ($m^2\ s^{-3}$) | $<[K_z]>$ ($m^2\ s^{-3}$) | $<[N]>$ ($s^{-1}$) |
|---|---|---|---|
| 3 total | $4\pm3\times10^{-9}$ | $2.7\pm1.2\times10^{-3}$ | $7.5\pm1\times10^{-4}$ |
| 3 <day 425 | $1.1\pm0.6\times10^{-9}$ | $2.7\pm1.2\times10^{-3}$ | $7.6\pm1\times10^{-4}$ |
| 5 early winter N-wind | $6\pm4\times10^{-10}$ | $2.0\pm1\times10^{-4}$ | $7.6\pm1\times10^{-4}$ |
| 7 late winter cold influx | $6\pm4\times10^{-8}$ | $3\pm1.2\times10^{-2}$ | $6.3\pm1\times10^{-4}$ |
| 8 late spring internal waves | $7\pm4\times10^{-11}$ | $3\pm1.5\times10^{-5}$ | $6.8\pm1\times10^{-4}$ |
| 9 late winter inertial waves | $3\pm2\times10^{-9}$ | $1.3\pm0.7\times10^{-3}$ | $7.6\pm1\times10^{-4}$ |

*References*

Amadori, M., S. Piccolroaz, L. Giovannini, D. Zardi, and M. Toffolon. 2018. Wind variability and Earth's rotation as drivers of transport in a deep, elongated subalpine lake: the case of Lake Garda. J. Limnol. **77**: 505-521.

Ambrosetti, W., L. Barbanti. 1999. Deep water warming in lakes: an indicator of climatic change. J. Limnol. **58**: 1-9.

Antenucci, J., J. Imberger. 2003. The Seasonal Evolution of Wind/Internal Wave Resonance in Lake Kineret. Limnol. Oceanogr. **48**: 2055-2061.

Berger, S. A., S. Diehl, H. Stibor, G. Trommer, M. Ruhenstroth, A. Wild, A. Weigert, C. Gerald Jäger, and M. Striebel. 2007. Water temperature and mixing depth affect timing and magnitude of events during spring succession of the plankton. Oecologia **150**: 643-654.

Boegman, L., J. Imberger, G. N. Ivey, and J. P. Antenucci. 2003. High-frequency internal waves in large stratified lakes. Limnol. Oceanogr. **46**: 895-919.

Boehrer, B., R. Fukuyama, and K. Chikita. 2008. Stratification of very deep, thermally stratified lakes. Geophys. Res. Lett. **35**: L16405. doi:10.1029/2008GL034519

Bouffard, D., and L. Boegman. 2013. A diapycnal diffusivity model for stratified environmental flows. Dyn. Atmos. Oc. **61-62**: 14-34.

Chalamalla, V. K., and S. Sarkar. 2015. Mixing, dissipation rate, and their overturn-based estimates in a near-bottom turbulent flow driven by internal tides. J. Phys. Oceanogr. **45**: 1969-1983.

Dauxois, T., A. Didier, and E. Falcon. 2004. Observations of near-critical reflection of internal waves in a stably stratified fluid. Phys. Fluids **16**: 1936-1941.

Dillon, T. M. 1982. Vertical overturns: a comparison of Thorpe and Ozmidov length scales. J. Geophys. Res. **87**: 9601-9613.

Dokulil, M. T. 2014. Impact of climate warming on European inland waters. Inland Wat. **4**: 27-40.

Ekman V.W. 1905. On the influence of the Earth's rotation on ocean-currents. Ark Math Astron Fys **2(11)**:1-52.

Eriksen, C. C. 1982. Observations of internal wave reflection off sloping bottoms. J. Geophys. Res. **87**: 525-538.

Farmer, D.M. 1978. Observations of long nonlinear internal waves in a lake. J. Phys. Oceanogr. **8**: 63-73.

Fer, I., U. Lemmin, and S. A. Thorpe. 2002. Winter cascading of cold water in Lake Geneva. J. Geophys. Res. **107**: C6. doi:10.1029/2001JC000828

Galbraith, P. S., and D. E. Kelley. 1996. Identifying overturns in CTD profiles. J. Atmos. Ocean. Tech. **13**: 688-702.

Garanaik, A., and S. K. Venayagamoorthy. 2019. On the inference of the state of turbulence and mixing efficiency in stably stratified flows. J. Fluid Mech. **867**: 323-333.

Garrett, C. J. R., and W. H. Munk. 1972. Space-time scales of internal waves. Geophys. Fluid Dyn. **3**: 225-264.

Gill, A. E. 1982. Atmosphere-Ocean Dynamics. Academic Press.

Giovannini, L., L. Laiti, D. Zardi, and M. de Franceschi. 2015. Climatological characteristics of the Ora del Garda wind in the Alps. Int. J. Clim. **35**: 4103-4115.

Goudsmit, G.-H., F. Peeters, M. Gloor, and A. Wüest. 1997. Boundary versus internal diapycnal mixing in stratified natural waters. J. Geophys. Res. **102**: 27903-27914. doi:10.1029/97JC01861

Goudsmit, G.-H., H. Burchard, F. Peeters, and A. Wüest. 2002. Application of k-ϵ turbulence models to enclosed basins: The role of internal seiches. J. Geophys. Res. **107**: 3230. doi:10.1029/2001JC000954

Gregg, M.C. 1989. Scaling turbulent dissipation in the thermocline. J. Geophys. Res. **94**: 9686-9698.

Gregg, M. C., E. A. D'Asaro, J. J. Riley, and E. Kunze. 2018. Mixing efficiency in the ocean. Annu. Rev. Mar. Sci. **10**: 443-473.

Imboden, D. M., and A. Wüest. 1995. Mixing mechanisms in lakes. In Lerman, A., D. M. Imboden, and J.R. Gat (editors), "Physics and chemistry of lakes", Springer.

IOC, SCOR, IAPSO. 2010. The international thermodynamic equation of seawater – 2010: Calculation and use of thermodynamic properties. Intergovernmental Oceanographic Commission, Manuals and Guides No. 56, UNESCO.

LeBlond, P. H., and L. A. Mysak. 1978. Waves in the Ocean. Elsevier.

Lemmin, U., C. H. Mortimer, and E. Bäuerle. 2005. Internal seiche dynamics in Lake Geneva. Limnol. Oceanogr. **50**: 207-216.

Leoni, B., L. Garibaldi, and R. Gulati. 2014. How does interannual trophic variability caused by vertical water mixing affect reproduction and population density of the Daphnia longispina group in Lake Iseo, a deep stratified lake in Italy. Inland Wat. **4**:193-203.

Li, S., and H. Li. 2006. Parallel AMR code for compressible MHD and HD equations. T-7, MS B284, Theoretical division, Los Alamos National Laboratory, http://math.lanl.gov/Research/Highlights/amrmhd.shtml

Lorke, A., F. Peeters, and E. Bäuerle. 2006. High-frequency internal waves in the littoral zone of a large lake. Limnol. Oceanogr. **51**: 1935-1939.

Lorke, A. 2007. Boundary mixing in the thermocline of a large lake. J. Geophys. Res. **112**: C09019. doi:10.1029/2006JC004008

Lorrai, C., L. Umlauf, J. K. Becherer, A. Lorke, and A. Wüest. 2011. Boundary mixing in lakes: 2. Combined effects of shear-and convectively induced turbulence on basin-scale mixing. J. Geophys. Res. **116**: C10018. doi:10.1029/2011JC007121

Mater, B. D., S. K. Venayagamoorthy, L. St. Laurent and J. N. Moum. 2015. Biases in Thorpe scale estimation of turbulence dissipation. Part I: Assessments from large-scale overturns in oceanographic data. J. Phys. Oceanogr. **45**: 2497-2521.

Matsumoto, and Y. Hoshino. 2004. Onset of turbulence by a Kelvin-Helmholtz vortex. Geophys. Res. Lett. **31**: L02807. doi:10.1029/2003GL018195.

Oakey, N. S. 1982. Determination of the rate of dissipation of turbulent energy from simultaneous temperature and velocity shear microstructure measurements. J. Phys. Oceanogr. **12**: 256-271.

Osborn, T. R. 1980. Estimates of the local rate of vertical diffusion from dissipation measurements. J. Phys. Oceanogr. **10**: 83-89.

Perroud, M., S. Goyette, A. Martynov, M. Beniston, and O. Anneville. 2009. Simulation of multiannual thermal profiles in deep Lake Geneva: A comparison of one-dimensional lake models. Limnol. Oceanogr. **54**: 1574-594.

Phillips, O. M. 1971. On spectra measured in an undulating layered medium. J. Phys. Oceanogr. **1**: 1-6.

Piccolroaz, S., M. Amadori, M. Toffolon, and H. A. Dijkstra. 2019. Importance of planetary rotation for ventilation processes in deep elongated lakes: Evidence from Lake Garda (Italy). Sci. Rep. **9**: 8290.

Polzin K. L., J. M. Toole, J. R. Ledwell, and R.W. Schmitt. 1997. Spatial variability of turbulent mixing in the abyssal ocean. Science **276**: 93-96.

Portwood, G. D., S. M. de Bruyn Kops, and C. P. Caulfield. 2019. Asymptotic dynamics of high dynamic range stratified turbulence. Phys. Rev. Lett. **122**: 194504.

Preusse, M. 2012. Properties of internal solitary waves in deep temperate lakes. Ph.D-thesis Univ. Konstanz.

Ravens, T. M., O. Kocsis, A. Wüest and N. Granin. 2000. Small-scale turbulence and vertical mixing in Lake Baikal. Limnol. Oceanogr. **45**: 159-173.

Salmaso, and F. Decet. 1998. Interactions of physical, chemical and biological processes affecting the seasonality of mineral composition and nutrient cycling in the water column of a deep subalpine lake (Lake Garda, Northern Italy). Arch. Hydrobiol. **142**: 385-414.

Salmaso, N., G. Morabito, R. Modello, L. Garibaldi, M. Simona, F. Buzzi, and D. Ruggiu. 2003. A synoptic study of phytoplankton in the deep lakes south of the Alps (lakes Garda, Iseo, Como, Lugano, and Maggiore). J. Limnol. **62**: 207-227.

Salmaso, N. 2005. Effects of climatic fluctuations and vertical mixing on the interannual trophic variability of Lake Garda, Italy. Limnol. Oceanogr. **50**: 553-565.

Sarkar, S., and A. Scotti. 2017. From topographic internal gravity waves to turbulence. Ann. Rev. Fluid Mech. **49**: 195-220.

Tennekes, H., and J. L. Lumley. 1972. A first in Turbulence. MIT Press.

Thorpe, S. A. 1977. Turbulence and mixing in a Scottish loch. Phil. Trans. Roy. Soc. Lond. A **286**: 125-181.

Thorpe, S. A., J. M. Keen, R. Jiang, and U. Lemmin. 1996. High frequencu internal waves in Lake Geneva. Phil. Trans. R. Soc. Lond. A **354**: 237-257.

Toffolon M., S. Piccolroaz, and H. A. Dijkstra. 2017. A plunge into the depths of Italy's Lake Garda. Eos **98**. doi:10.1029/2017EO074499

Valerio, G., M. Pilotti, M. Clelia, and J. Imberger. 2012. The structure of basin scale internal waves in a stratified lake in response to lake bathymetry and wind spatial and temporal distribution: Lake Iseo, Italy. Limnol. Oceanogr. **57**: 772-786.

Valerio, G., M. Pilotti, M. Lau, and M. Hupfer. 2019. Oxycline oscillations induced by internal waves in deep Lake Iseo. Hydrol. Earth Syst. Sci. **23**: 1763-1777.

van Haren, H. 2017. Exploring the vertical extent of breaking internal wave turbulence above deep-sea topography. Dyn. Atmos. Oc. **77**: 89-99.

van Haren, H. 2018. Philosophy and application of high-resolution temperature sensors for stratified waters. Sensors **18**: 3184. doi:10.3390/s18103184

van Haren, H. 2019. Open-ocean interior moored sensor turbulence estimates, below a Meddy. Deep-Sea Res. I **144**: 75-84.

van Haren, H., and L. Gostiaux. 2012. Detailed internal wave mixing above a deep-ocean slope. J. Mar. Res. **70**: 173-197.

van Haren, H., L. Maas, J. T. F. Zimmerman, H. Ridderinkhof, and H. Malschaert. 1999. Strong inertial currents and marginal internal wave stability in the central North Sea. Geophys. Res. Lett. **26**: 2993-2996.

van Haren, H., A. A. Cimatoribus, and L. Gostiaux. 2015. Where large deep-ocean waves break. Geophys. Res. Lett. **42**: 2351-2357. doi:10.1002/2015GL063329

Wang, Y., C. Hutter, and E. Bäuerle. 2000. Wind-induced baroclinic response of Lake Constance. Ann. Geophys. 18: 1488-1501.

Warhaft, Z. 2000. Passive scalars in turbulent flows. Ann. Rev. Fluid Mech. **32**: 203-240.

Wetzel, R. G. 2001. Limnology: Lake and River Ecosystems, 3rd ed. Academic Press, San Diego.

Winters, K. B. 2015. Tidally driven mixing and dissipation in the boundary layer above steep submarine topography. Geophys. Res. Lett. **42**: 7123-7130. doi:10.1002/2015GL064676

Wüest, A., and A. Lorke. 2003. Small-scale hydrodynamics in lakes. Ann. Rev. Fluid Mech. **35**: 373-412.

**Fig. 1**. (**a**) Lake Garda (I) bathymetry, with sub-surface mooring site (red star) and local horizontal coordinate system in insert. The red and yellow dots indicate CTD and meteorological stations, respectively. (**b**) The mooring top-buoy was towed to the site from Brenzone at the eastern shore using a fast-boat of the firefighters nautical rescue team of the Autonomous Province of Trento.

**Fig. 2**. Shipborne CTD-observations near the mooring location recorded on 14 November 2017 (black) and 13 March 2018 (blue; 10 days after cold water influx). (**a**) Potential temperature. (**b**) Salinity (note the small x-axis scale). (**c**) Density anomaly with pressure corrections relative to the surface level. (**d**) Potential temperature-density anomaly relationship, from the portion of CTD-data in a.-c. between 190 and 334 m which includes the range of moored temperature sensors.

**Fig. 3**. Overview of about 300 days of observations in and near Lake Garda. (**a**) Wind components measured at the station near the northern end of the lake and plotted in oceanographic convention (facing with the wind), with along-lake component in blue and (the weaker) cross-lake component in green. (**b**) Current-flow components measured at the mooring near mid-depth, with along-lake component in black and cross-lake component in green. (**c**) Air temperature measured at the northern end of the lake (blue) and water temperature near mid-depth z =-191 m (red). (**d**) Time-depth series of potential temperature from high-resolution temperature sensors, with missing sensors interpolated (see text). The lake-floor is at the horizontal axis. (**e**) Logarithm of vertically lower ~50-m (black) and upper ~100-m (green) averaged turbulence dissipation rates.

**Fig. 4**. 275 day (covering yeardays 150 to 425) averaged spectra. (**a**) Data from the mid-depth (z = -187 m) current meter, with arbitrary shifts for the vertical to compare different parameters. The spikes in T (and p) at frequencies >10 cpd (short for cycles per day) are due to low resolution of the instrumental sampling. KE denotes the total kinetic energy and $W_y$ the along-

lake wind spectrum. Several harmonic tidal frequencies are indicated (diurnal solar $S_1$, semidiurnal $S_2$, fourth-diurnal $S_4$), besides inertial frequency f, 1.1f (magenta line) and its first harmonic 2f, mean local large-scale buoyancy frequency N and maximum local small-scale buoyancy frequency $N_{max}$. (**b**) Data from five temperature sensors at indicated z-levels. The spectra are scaled with turbulence inertial subrange $\sigma^{-5/3}$ (of which the slope is indicated by horizontal black bar). The slope of $\sigma^{-1/3}$ (dashed black line) indicates either a non-inertial turbulence subrange or the canonical internal wave slope of $\sigma^{-2}$ in an unscaled spectrum. The small horizontal bar indicates the variation in N.

**Fig. 5**. Four days detail from Fig. 3 (12-16 November 2017) on moderate wintertime wind/internal wave mixing. (**a**) Along-lake wind components measured at the station near the northern end of the lake (blue) and near the southern end (light-blue), plotted in oceanographic convention. (**b**) Current-flow components measured at the mooring near mid-depth, with along-lake component in black and cross-lake component in green. (**c**) Time-depth image of potential temperature. Contours in black are drawn every 0.018°C. In the blue part the weak stratification yields approximately N ≈ f. The white bar indicates the inertial period, the light-blue bar the mean buoyancy period and the purple bar the minimum buoyancy period. Tick-marks are at semidiurnal periods. (**d**) Logarithm of dissipation rate per vertical increment and per time step, with θ-contours from c. Red ellipses highlight near-bottom mixing to the left and interior internal wave breaking in the centre.

**Fig. 6**. Magnifications of potential temperature from Fig. 5c on day 318 (14 November 2017). (**a**) 8-h example period. The green bar indicates the period of subplot b. (**b**) 4-h period. The green bar indicates the period of subplot c., the light-blue bar indicates the mean buoyancy period. (**c**) 1-h period, with focus on the lower half of the observed range and with a different color scale but identical contour interval of 0.018°C. The purple bar indicates the minimum buoyancy period. The ellipses indicate several overturning events.

**Fig. 7**. As Fig. 5 with the same scales, but for late winter (3-7 March 2018) cold water influx and strong overturning.

**Fig. 8**. As Fig. 5 with the same scales, but for summertime (8-12 June 2017) weak interior and moderate near-bottom mixing induced by semidiurnal period internal waves.

**Fig. 9**. As Fig. 5 with the same scales, but for a ten-day period of late winter (17-27 March 2018) inertial wave (interior) mixing.

**Fig. 10**. Vertical profiles of (logarithms of) time-mean turbulence values from the moored temperature data of Fig. 5 (early winter northerly winds, black), Fig. 7 (late winter cold water influx, blue), Fig. 8 (late spring/early summer internal waves, light-blue), Fig. 9 (late winter near-inertial waves, purple). (**a**) Dissipation rate. (**b**) Eddy diffusivity, after averaging flux first. (**c**) Buoyancy frequency from reordered profiles.

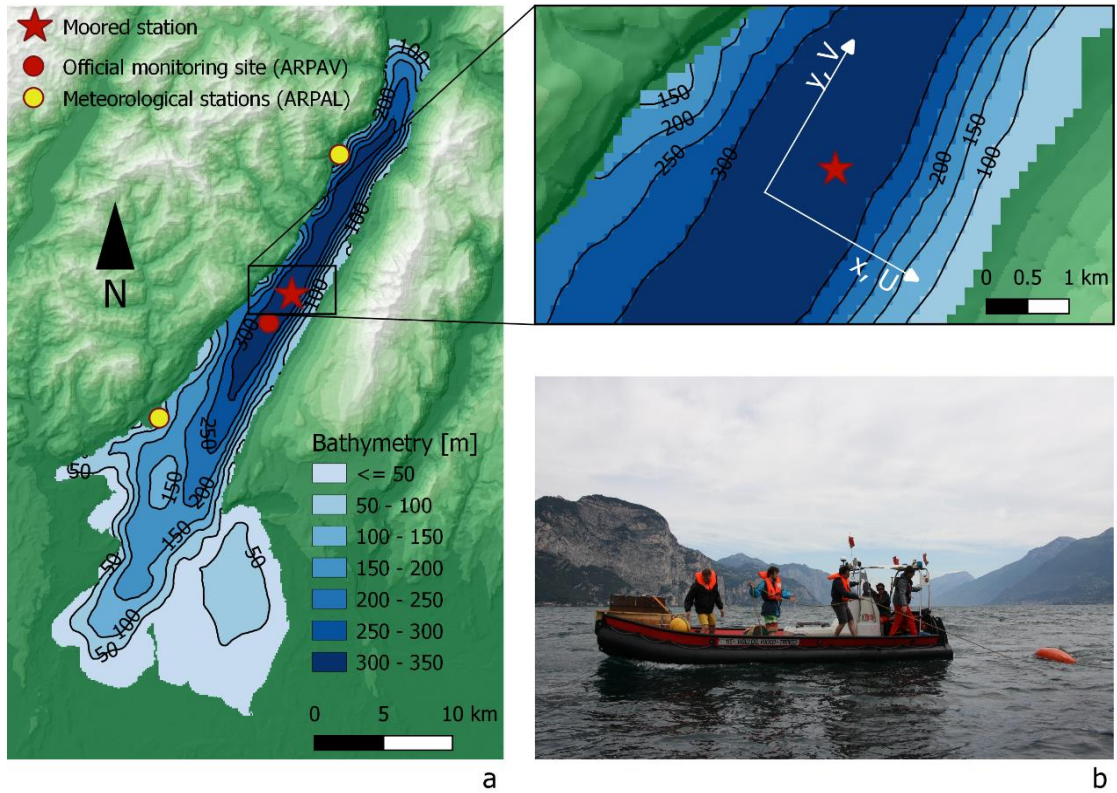

**Fig. 1**. (**a**) Lake Garda (I) bathymetry, with sub-surface mooring site (red star) and local horizontal coordinate system in insert. The red and yellow dots indicate CTD and meteorological stations, respectively. (**b**) The mooring top-buoy was towed to the site from Brenzone at the eastern shore using a fast-boat of the firefighters nautical rescue team of the Autonomous Province of Trento.

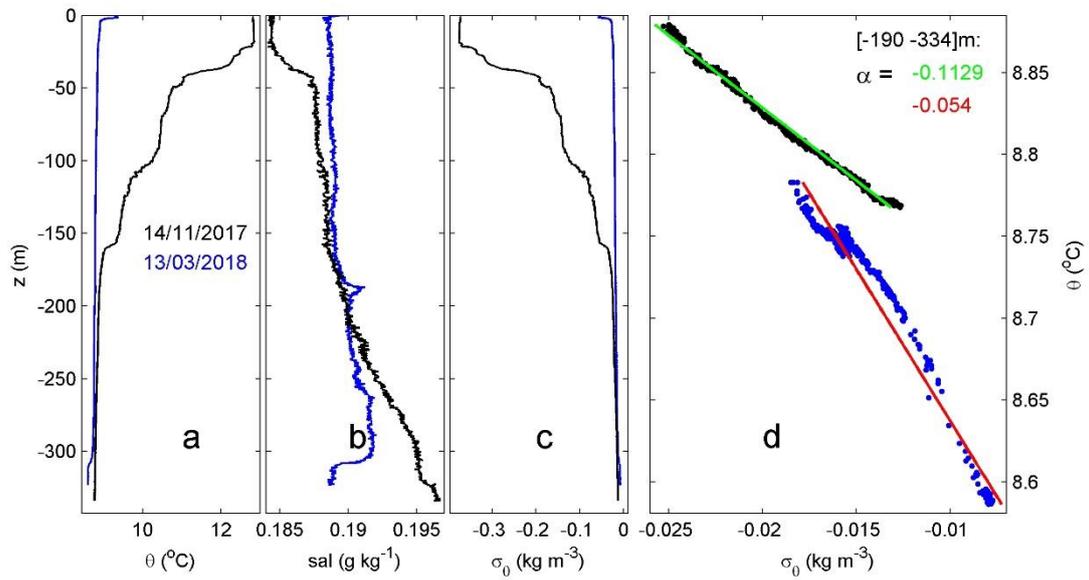

**Fig. 2**. Shipborne CTD-observations near the mooring location recorded on 14 November 2017 (black) and 13 March 2018 (blue; 10 days after cold water influx). (**a**) Potential temperature. (**b**) Salinity (note the small x-axis scale). (**c**) Density anomaly with pressure corrections relative to the surface level. (**d**) Potential temperature-density anomaly relationship, from the portion of CTD-data in a.-c. between 190 and 334 m which includes the range of moored temperature sensors.

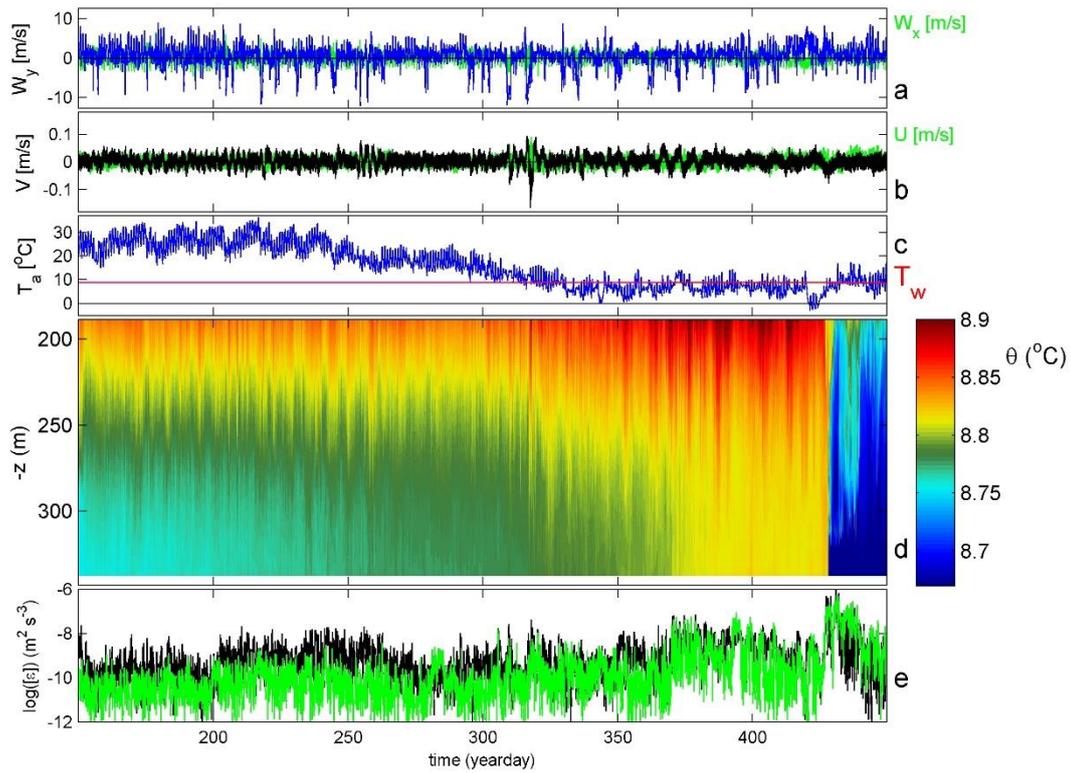

**Fig. 3**. Overview of about 300 days of observations in and near Lake Garda. (**a**) Wind components measured at the station near the northern end of the lake and plotted in oceanographic convention (facing with the wind), with along-lake component in blue and (the weaker) cross-lake component in green. (**b**) Current-flow components measured at the mooring near mid-depth, with along-lake component in black and cross-lake component in green. (**c**) Air temperature measured at the northern end of the lake (blue) and water temperature near mid-depth z = -191 m (red). (**d**) Time-depth series of potential temperature from high-resolution temperature sensors, with missing sensors interpolated (see text). The lake-floor is at the horizontal axis. (**e**) Logarithm of vertically lower ~50-m (black) and upper ~100-m (green) averaged turbulence dissipation rates.

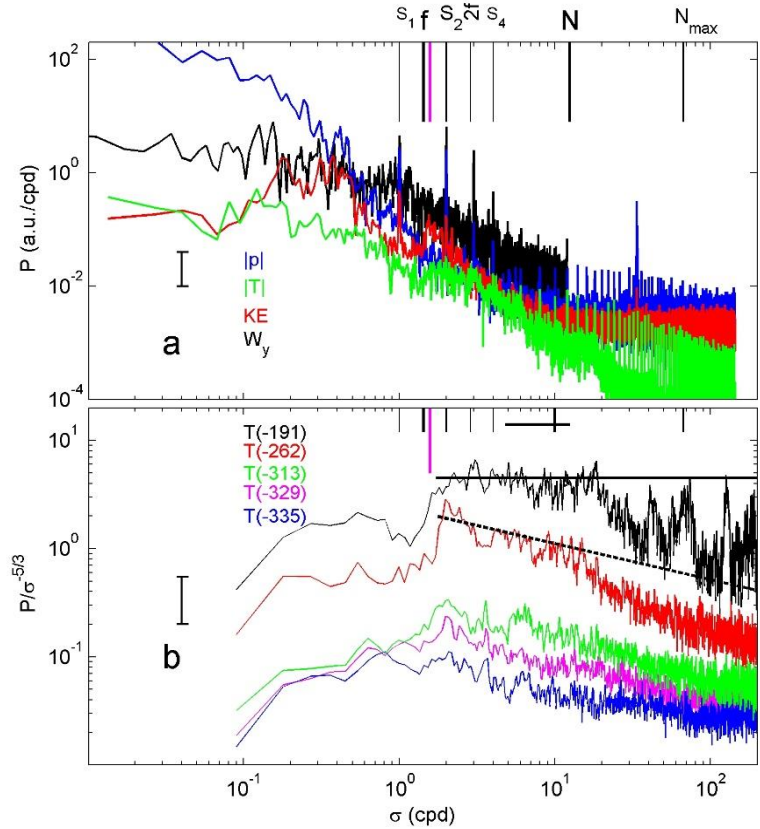

**Fig. 4**. 275 day (covering yeardays 150 to 425) averaged spectra. (**a**) Data from the mid-depth (z = -187 m) current meter, with arbitrary shifts for the vertical to compare different parameters. The spikes in T (and p) at frequencies >10 cpd (short for cycles per day) are due to low resolution of the instrumental sampling. KE denotes the total kinetic energy and $W_y$ the along-lake wind spectrum. Several harmonic tidal frequencies are indicated (diurnal solar $S_1$, semidiurnal $S_2$, fourth-diurnal $S_4$), besides inertial frequency f, 1.1f (magenta line) and its first harmonic 2f, mean local large-scale buoyancy frequency N and maximum local small-scale buoyancy frequency $N_{max}$. (**b**) Data from five temperature sensors at indicated z-levels. The spectra are scaled with turbulence inertial subrange $\sigma^{-5/3}$ (of which the slope is indicated by horizontal black bar). The slope of $\sigma^{-1/3}$ (dashed black line) indicates either a non-inertial turbulence subrange or the canonical internal wave slope of $\sigma^{-2}$ in an unscaled spectrum. The small horizontal bar indicates the variation in N.

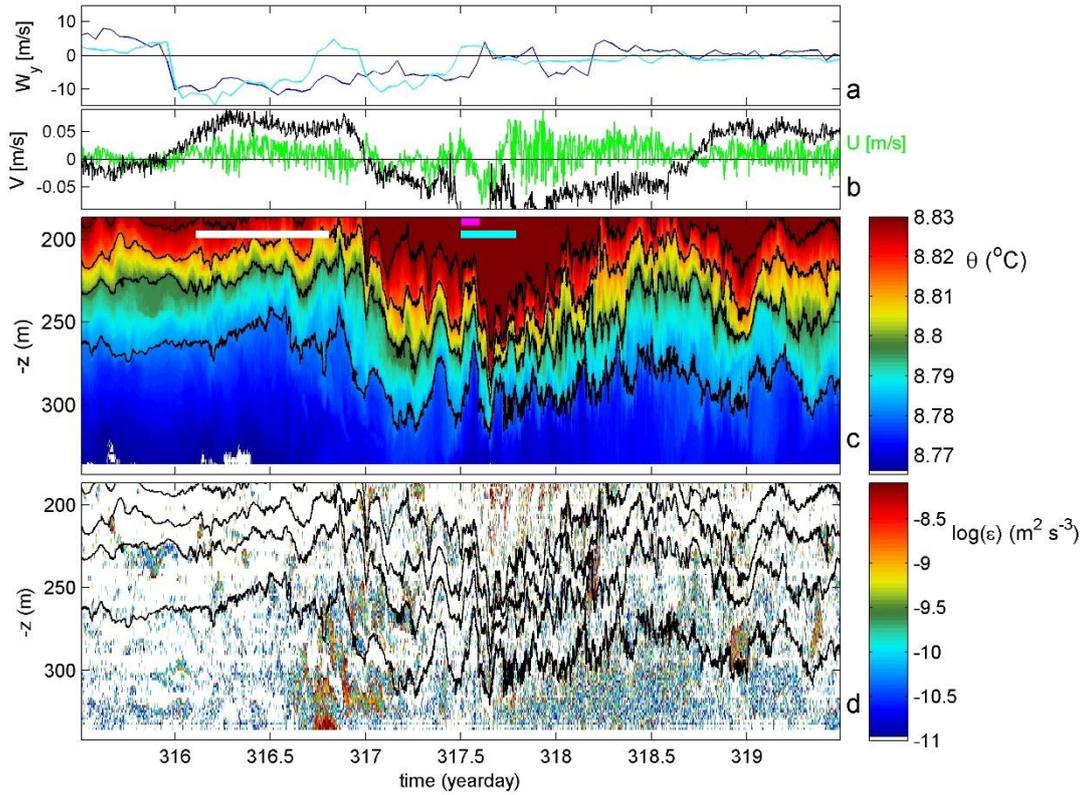

**Fig. 5**. Four days detail from Fig. 3 (12-16 November 2017) on moderate wintertime wind/internal wave mixing. (**a**) Along-lake wind components measured at the station near the northern end of the lake (blue) and near the southern end (light-blue), plotted in oceanographic convention. (**b**) Current-flow components measured at the mooring near mid-depth, with along-lake component in black and cross-lake component in green. (**c**) Time-depth image of potential temperature. Contours in black are drawn every 0.018°C. In the blue part the weak stratification yields approximately N ≈ f. The white bar indicates the inertial period, the light-blue bar the mean buoyancy period and the purple bar the minimum buoyancy period. Tick-marks are at semidiurnal periods. (**d**) Logarithm of dissipation rate per vertical increment and per time step, with θ-contours from c. Red ellipses highlight near-bottom mixing to the left and interior internal wave breaking in the centre.

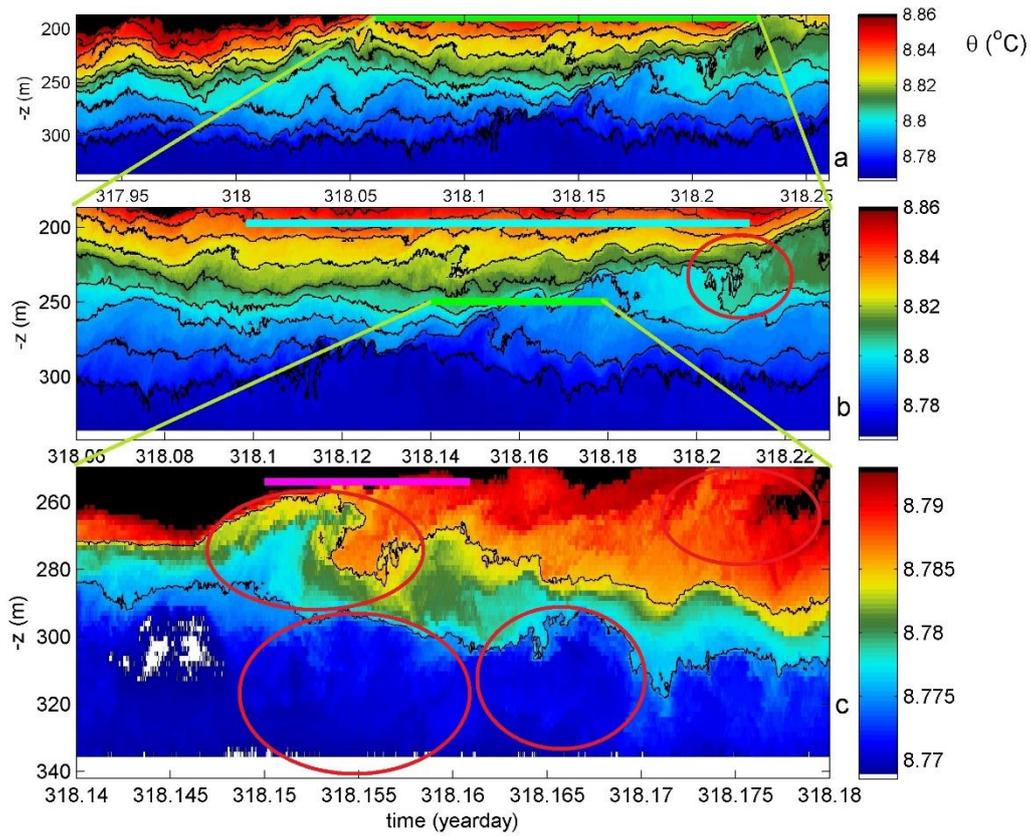

**Fig. 6**. Magnifications of potential temperature from Fig. 5c on day 318 (14 November 2017). (**a**) 8-h example period. The green bar indicates the period of subplot b. (**b**) 4-h period. The green bar indicates the period of subplot c., the light-blue bar indicates the mean buoyancy period. (**c**) 1-h period, with focus on the lower half of the observed range and with a different color scale but identical contour interval of 0.018°C. The purple bar indicates the minimum buoyancy period. The ellipses indicate several overturning events.

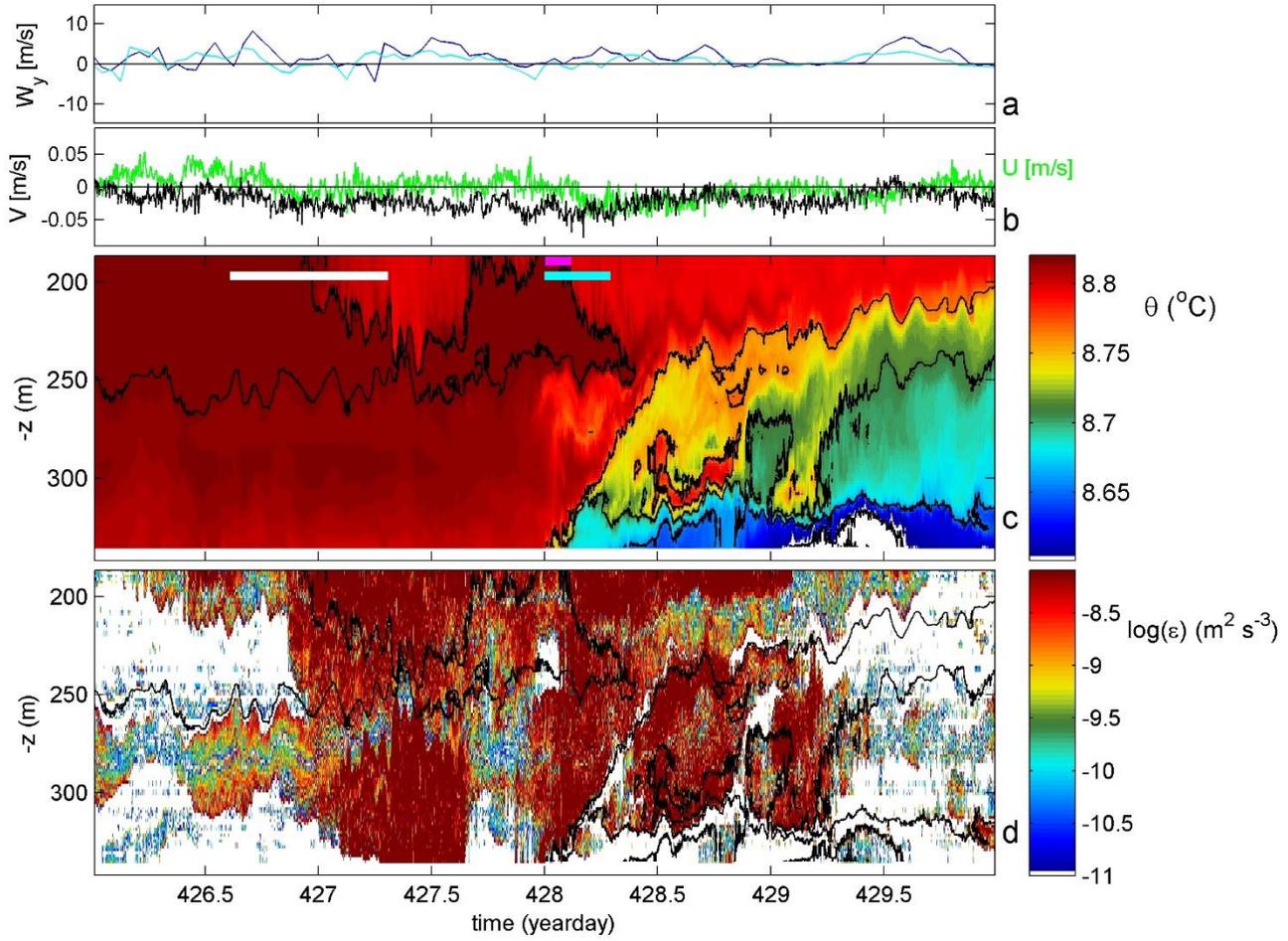

**Fig. 7**. As Fig. 5 with the same scales, but for late winter (3-7 March 2018) cold water influx and strong overturning.

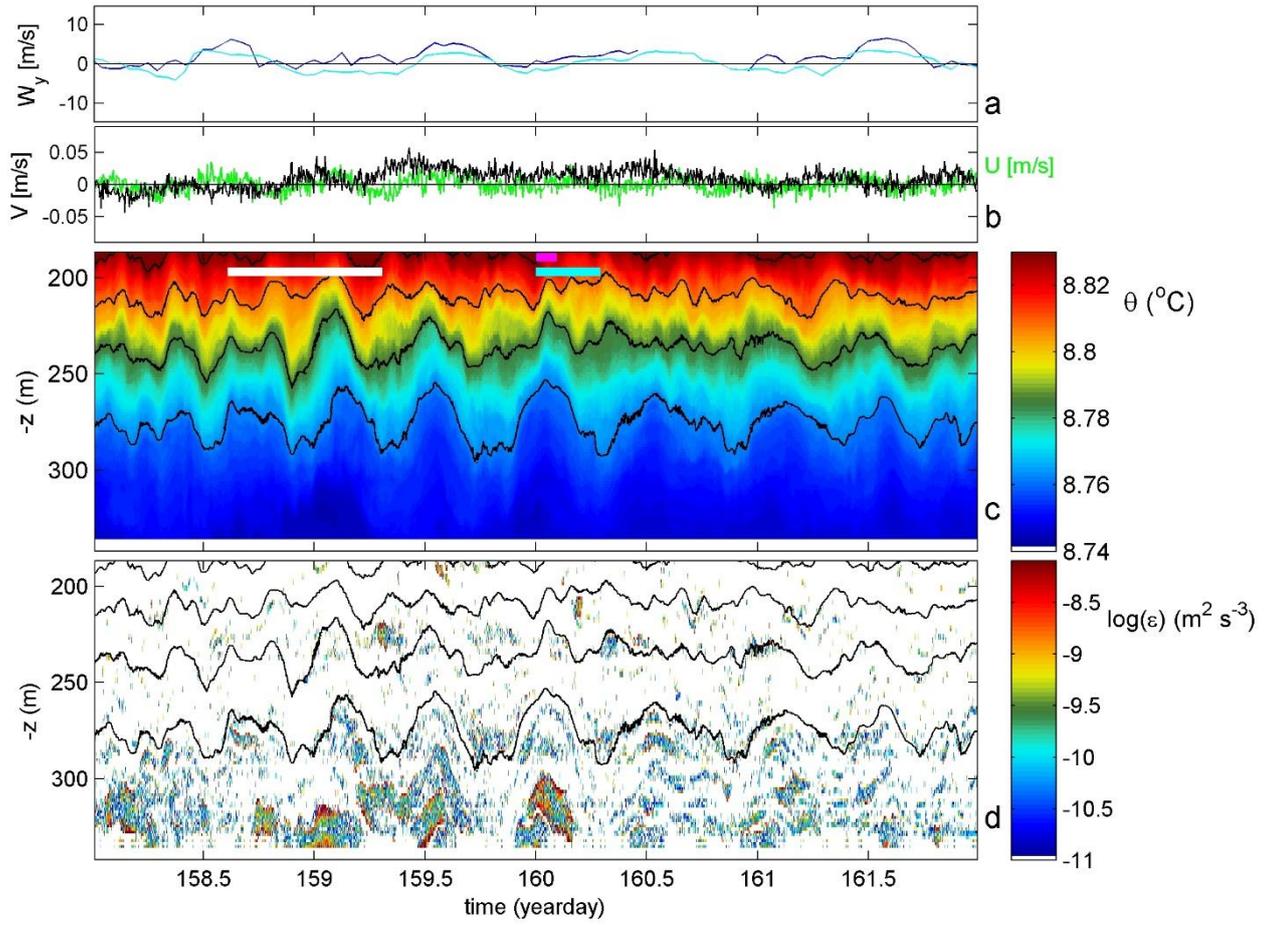

**Fig. 8**. As Fig. 5 with the same scales, but for summertime (8-12 June 2017) weak interior and moderate near-bottom mixing induced by semidiurnal period internal waves.

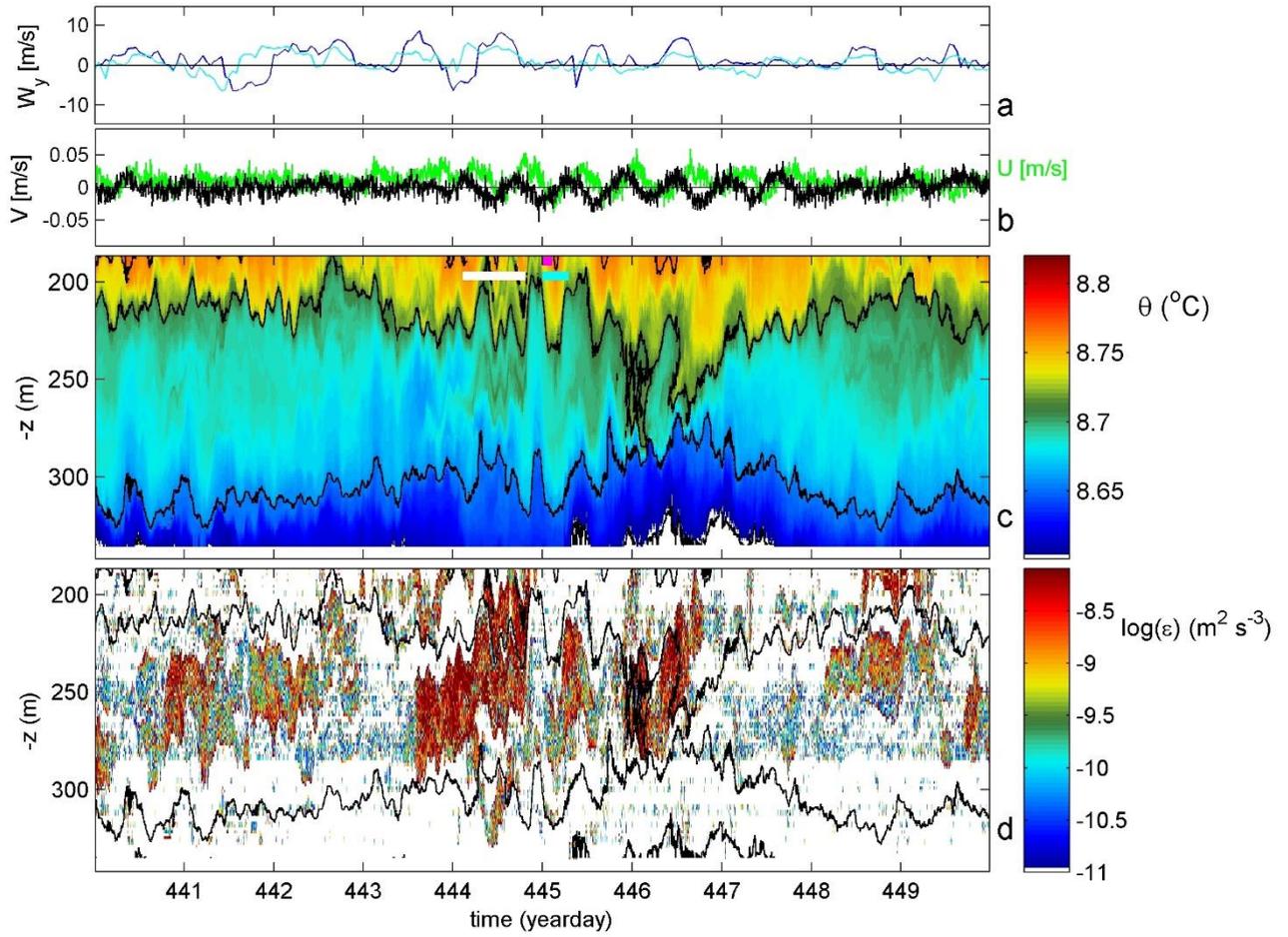

**Fig. 9**. As Fig. 5 with the same scales, but for a ten-day period of late winter (17-27 March 2018) inertial wave (interior) mixing.

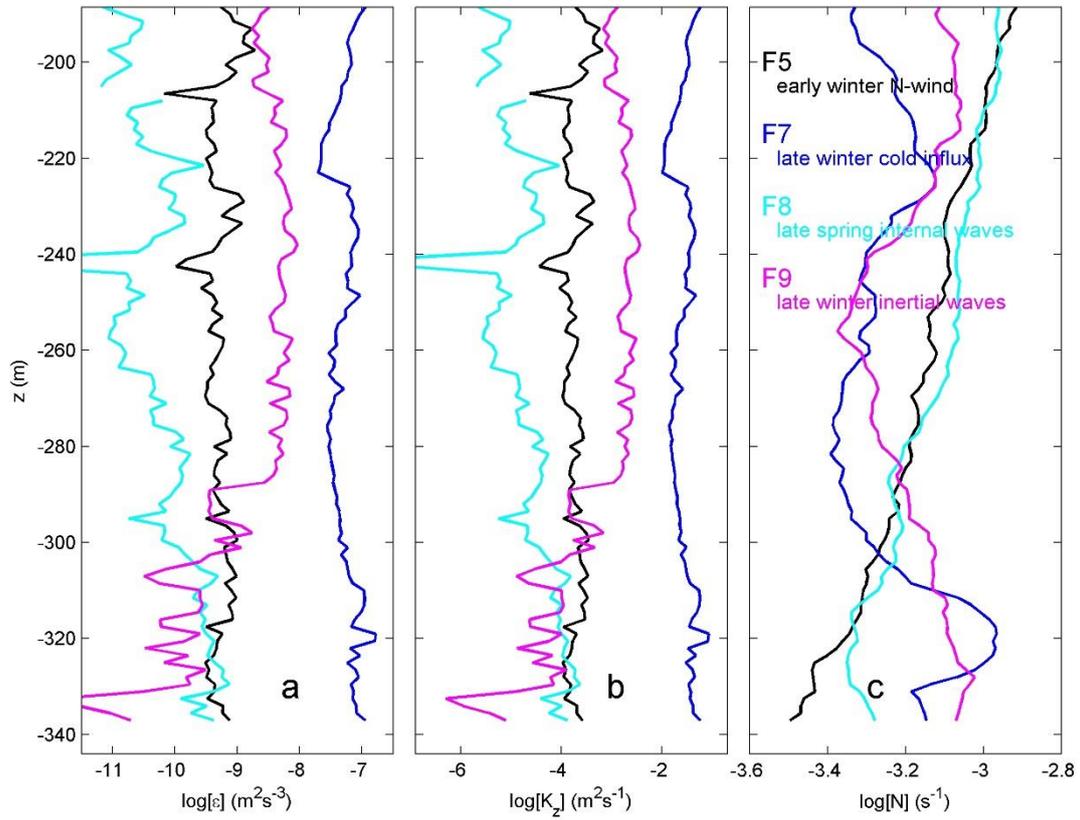

**Fig. 10**. Vertical profiles of (logarithms of) time-mean turbulence values from the moored temperature data of Fig. 5 (early winter northerly winds, black), Fig. 7 (late winter cold water influx, blue), Fig. 8 (late spring/early summer internal waves, light-blue), Fig. 9 (late winter near-inertial waves, purple). (**a**) Dissipation rate. (**b**) Eddy diffusivity, after averaging flux first. (**c**) Buoyancy frequency from reordered profiles.